\documentclass[preprint2]{aastex}

\def\wig#1{\mathrel{\hbox{\hbox to 0pt{%
          \lower.5ex\hbox{$\sim$}\hss}\raise.4ex\hbox{$#1$}}}}

\slugcomment{Accepted for publication in ApJ, October 2006}

\shorttitle{}
\shortauthors{Leggett et al.}

\begin{document}

\title{3.6--7.9$\,\mu$m Photometry of L and T Dwarfs and the
Prevalence of Vertical Mixing in their Atmospheres}

\author{S. K. Leggett}
\affil{Gemini Observatory,  670 N. A'ohoku Place Hilo HI 96720}
\email{sleggett@gemini.edu}

\author{D. Saumon}
\affil{Los Alamos National Laboratory, MS P365, Los Alamos, NM 87545}

\author{M. S. Marley}
\affil{MS 245-3, NASA Ames Research Center, Moffett Field, CA 94035}

\author{T. R. Geballe}
\affil{Gemini Observatory,  670 N. A'ohoku Place, Hilo HI 96720}

\author{D. A. Golimowski and D. Stephens}
\affil{Department of Physics \& Astronomy, Johns Hopkins University, 
3400 North Charles Street, Baltimore, MD 21218}

\and 

\author{X. Fan}
\affil{Steward Observatory, 933 N. Cherry Avenue, Tucson, AZ 85721}

\begin{abstract}

We present new $L'$ (3.75~$\mu$m) photometry of six L and T
dwarfs, and $M'$ (4.70~$\mu$m) photometry of ten L and T dwarfs,
observed at Gemini (North) Observatory, and new 3.55, 4.49, 5.73 and
7.87~$\mu$m photometry of nine L and T dwarfs, obtained with the
{\it Spitzer Space Telescope}. The sample includes unusually blue and red
dwarfs selected from our near-infrared studies. The data are combined
with previously published $L'$, $M'$ and {\it Spitzer}
photometry of L and T dwarfs, and trends of colors with spectral type and
other colors are examined.  Recent model atmospheres by Marley and Saumon
are used to generate synthetic colors for ranges of effective
temperature, gravity, grain sedimentation efficiency, metallicity and vertical
mixing efficiency.  We explore how these parameters affect the mid-infrared colors
of L and T dwarfs and find that the data are modelled satisfactorily 
only if substantial vertical mixing occurs in both L- and T-dwarf
atmospheres.  The location and range of the L and T dwarf sequences in 
IRAC color-color and color-magnitude diagrams is also only reproduced if this mixing occurs,
with a range of efficiency described by $K_{zz} \sim 10^2$--$10^6\,$cm$^2$~s$^{-1}$.
The colors of the unusually red dwarfs are best reproduced
by non-equilibrium models with low sedimentation efficiency, i.e. thick
cloud decks, and the colors of the unusually blue dwarfs by
non-equilibrium models with high sedimentation efficiency, i.e. thin
cloud decks. The $K$--$L'$ and {\it Spitzer} [3.55]--[4.49] 
colors can be used as indicators of effective temperature for L and T dwarfs, 
but care must be taken to include gravity and metallicity effects for late-T dwarfs and 
vertical mixing for both late-L and T dwarfs.

\end{abstract}


\section{Introduction}

The Sloan Digital Sky Survey (SDSS; York et al.\ 2000) and the Two Micron
All Sky Survey (2MASS; Beichman et al.\ 1998, Skrutskie et al.\ 2006) 
have revealed large numbers
of ultracool low-mass field dwarfs, known as L and T dwarfs.  The
effective temperatures ($T_{\rm eff}$) of L dwarfs are $\sim
1450$--2200~K, and those of currently known T dwarfs are $\sim
700$--1450~K (Golimowski et al.\ 2004a, hereafter G04; Vrba et al.\ 2004). 
 
As $T_{\rm eff}$ drops below that of the late-M dwarfs, the spectral 
energy distributions (SEDs) of ultracool dwarfs are dramatically changed
by chemical and cloud processes in their atmospheres.
First, iron and silicate grains condense high in the atmosphere and form
clouds that veil gaseous absorption bands and redden the near-infrared
colors of L dwarfs.  As $T_{\rm eff}$ falls, these clouds form progressively
deeper in the atmosphere and become more optically thick.  By 
$\sim 1500$--1700~K, the effect of the clouds on the emergent flux is greatest 
(Ackerman \& Marley 2001).  At lower $T_{\rm eff}$, the clouds lie near or below the base of
the wavelength-dependent photosphere and only marginally affect the
flux distributions of T dwarfs.  The rapidity, with respect to $T_{\rm eff}$, 
with which the clouds diminish is not well understood and several mechanisms have been proposed
(e.g. Burgasser et al.\ 2002b, Burrows, Sudarsky \& Hubeny\ 2006, Knapp et al.\ 2004, hereafter K04). 
G04 showed that $T_{\rm eff}$ is nearly constant ($\sim 1300$--1500~K) for spectral types L7 to T4.  
Thus, for these spectral types, the primary cause of the 
observed spectral change is apparently rapid alteration of the vertical properties and 
distribution of the cloud, over a small range in $T_{\rm eff}$.
The changing cloud properties alter the optical depths and brightness temperatures from 
which the flux emerges.  As the clouds depart, the upper atmosphere cools and   
CH$_4$ supplants CO as the dominant carbon-bearing molecule. CH$_4$  appears 
in the 3--4~$\mu$m spectra of mid-L dwarfs (Noll et al.\ 2000), the
2.2--2.5 and $8~\mu$m spectra of late-L dwarfs (Geballe et al.\ 2002;
Roellig et al.\ 2004; Cushing et al.\ 2006), and the $1.6~\mu$m spectra of
T0 dwarfs (Geballe et al.\ 2002). Together, increasing CH$_4$ absorption
and sinking cloud decks cause progressively bluer near-infrared colors
of T dwarfs. 

Over 500 ultracool field dwarfs have been confirmed and studied
in the near-infrared  (e.g. Burgasser et al.\ 2006b, 
Chiu et al.\ 2006 (hereafter C06), Kendall et al.\ 2004, Tinney et al.\ 2005).
Our group has so far obtained accurate
$J$, $H$, and $K$ photometry on the Mauna Kea Observatories (MKO) system (Simons \& Tokunaga
2002, Tokunaga et al.\ 2002) and $R \approx 500$, 0.9--2.5~$\mu$m
spectroscopy for 102 L dwarfs and 65 T dwarfs.  This sample is
summarized by K04 and C06.  Leggett et al. 
(2002) and G04 extended our study to longer wavelengths, presenting $L'$
(3.75~$\mu$m) and $M'$ (4.70~$\mu$m) photometry for a subset of
the $JHK$ sample.  In this paper, we extend our study further
by presenting broadband photometry for wavelengths up to 7.9~$\mu$m.

The 3--15~$\mu$m mid-infrared region is interesting for ultracool-dwarf studies
for many reasons.  This region contains molecular absorption bands of
CH$_4$, CO, H$_2$O and NH$_3$ (Cushing et al.\ 2006) that are prominent in 
the spectra of ultracool dwarfs at various temperatures. 
Figure~1 shows observed and synthetic 3--10~$\mu$m spectra for late-L and T
dwarfs and identifies the principal absorbing species.  The vibrotational line 
lists for CH$_4$, NH$_3$ and H$_2$O are more complete in the mid-infrared than the
near-infrared, so the uncertainties in the synthetic mid-infrared spectra 
are smaller than those in the near-infrared spectra.  
Also, the mid-infrared flux emerges from photospheric
regions that have a much smaller range of brightness temperatures than those giving
rise to the
near-infrared flux (Ackerman \& Marley 2001). Hence the mid-infrared
spectrum should be easier to interpret than the near-infrared spectrum. 
Furthermore, mid-infrared photometry is more sensitive than
near-infrared photometry to vertical mixing of trace species, which can result in
photospheric chemical abundances very different from equilibrium (Saumon
et al.\ 2003; G04).

Obtaining mid-infrared photometry and spectroscopy with ground-based
telescopes is difficult, due to the high and rapidly varying 
background at these wavelengths, and to the strong telluric absorption
bands of CH$_4$, CO, CO$_2$, H$_2$O, N$_2$O and O$_3$.  The 
{\it Spitzer Space Telescope} (Werner et al.\ 2003) has made possible
observations at these and longer wavelengths that were previously
impossible from the ground.  We are using $Spitzer$'s Infrared Array Camera 
(IRAC; Fazio et al.\ 2004) and Infrared Spectrograph (IRS; Houck et al.\ 2004) 
to obtain 3.6--7.9~$\mu$m photometry and 6--15~$\mu$m low-resolution spectra 
of late-L and T dwarfs.  In this paper, we present the IRAC results along
with ground-based $L'$ and $M'$ photometry obtained using the Near InfraRed 
Imager and spectrograph (NIRI; Hodapp et al.\ 2003) at the Gemini (North) 
Observatory.  The IRS spectra will be presented in a future paper.

\begin{figure} \includegraphics[angle=-90,scale=.3]{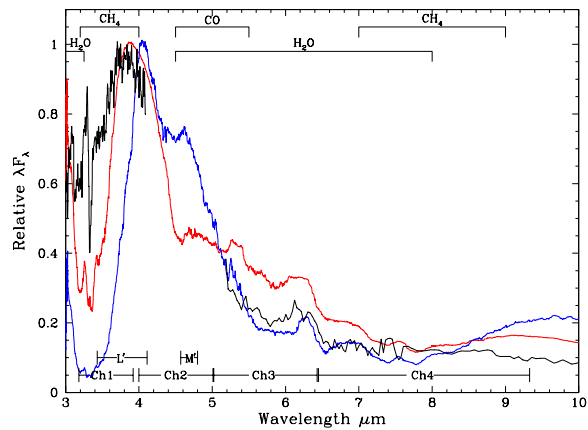} 
\caption{Observed and synthetic spectra of late-L and T dwarfs (normalized to
peak flux) in the wavelength interval where we report new photometry. The black 
curves are NIRI and IRS spectra of the L9 dwarf 2MASS J09083803+5032088
(Stephens private communication). The red curve is a synthetic spectrum for 
$T_{\rm eff}=1400$~K, $f_{\rm sed}=2$, and log $g=5$, representing late-L
types, and the blue curve is a synthetic spectrum for $T_{\rm eff}=$1000,
$f_{\rm sed}=4$, and log $g=5$, representing mid-T types (Marley \& Saumon,
private communication).  Strong molecular absorption bands are indicated, as are the bandpasses 
of the MKO $L^{\prime}$, MKO $M^{\prime}$ and IRAC channels 1--4 filters:
[3.55], [4.49], [5.73], [7.87].}
\end{figure}

\section{Observations}

\subsection{Gemini NIRI $L^{\prime}$ and $M^{\prime}$ Photometry}

We obtained MKO $L^{\prime}$ and/or $M^{\prime}$ images of 12 late-L and T 
dwarfs, listed in Table 1, on 10 dry and photometric nights between UT 2004 April 05 and 
UT 2005 January 28.  The data were obtained in queue-observing mode through programs
GN-2004A-Q-16 and GN-2004B-Q-43, using Gemini Observatory's NIRI.
The half-power bandpasses for the MKO $L'$ and $M'$ filters (Simons \& Tokunaga 
2002, Tokunaga et al.\ 2002) are shown in Figure~1 and listed in Table~2.  
Because the sky is bright at these wavelengths, 
we used NIRI's smallest pixel scale (0\farcs022 pix$^{-1}$) to avoid
saturating the detector.  The full array readout was used, providing a
field of view of 22\farcs5~$\times$~22\farcs5.  Given the small field of view, we acquired each
target using proper-motion-corrected coordinates (if available) and a
short $J$-band verification image.  During the $L'$ and $M'$ exposures,
the target was sequentially offset from the center of the detector toward 
each of the four corners, generating a four-position dither pattern with $6''$ 
offsets.  Offset guide stars were used.  Each $L^{\prime}$ image comprised
24 coadded exposures of 0.8~s, and each $M^{\prime}$ image comprised 34 coadded 
exposures of 0.5~s.  The total observation time per target was typically 1.5~hr, 
most of which was spent on the $M^{\prime}$ exposures.  Each set of science
images was accompanied by contemporaneous dark frames and images of bright 
photometric standard stars (Leggett et al.\ 2003) taken with the same filters.
Flat fields were generated by median-stacking the dithered science images.

The data were reduced using a version of the astronomical imaging pipeline 
ORAC-DR (Cavanagh et al.\ 2003) modified for NIRI.\footnote{The ORAC-DR Imaging
Data Reduction User Guide is available at 
http://www.starlink.ac.uk/star/docs/sun232.htx/sun232.html.}
The pipeline uses Starlink routines to perform
bad-pixel masking, dark subtraction, differencing of adjacent pairs,
flat-field creation and division, feature detection and matching between
object frames, and resampling.  The final combined image, termed a mosaic,
is generated with intensity offsets applied to give the most consistent 
results in the overlapping regions of the dithered images.  
Pairwise differencing allows more accurate sky subtraction and 
flat-fielding, although it yields 
positive and negative images of each source
in the mosaics.  The mosaics were binned $3\times3$ to enhance the visibility of
the L and T dwarfs.  The final pixel scale was therefore 0\farcs066~pix$^{-1}$.

We performed aperture photometry on both the positive and negative images of 
the L and T dwarfs using apertures of diameter $1''$--$2''$.  We defined the 
uncertainty to be the difference between the measurements of the positive and negative 
images.  This difference is similar to, or slightly larger than, the formal error calculated 
from the noise.  Table 1 lists the resulting $L^{\prime}$ and $M^{\prime}$ magnitudes,
as well as any previously published measurements.  Henceforth, we use the weighted averages
of the new and previously published measurements of all the dwarfs except 2MASS 
J15031961+2525196, whose previous uncertain $M'$ measurement is discarded in favor 
of the NIRI value.

\subsection{$Spitzer$ Four-Channel Photometry}

We obtained IRAC four-channel photometry (3.55, 4.49, 5.73, and 7.87~$\mu$m) of the 
nine L and T dwarfs listed in Table~3 between UT 2005 July 23 and UT 2006 July 06.
The data were obtained as part of 
our Cycle~2 {\it Spitzer} General Observer program \# 20514.  The half-power 
bandpasses for the four channels are listed in Table~2 and indicated in Figure~1.  
All four channels have 256$\times$256-pixel detectors with a pixel size 
of 1\farcs2$\times$1\farcs2, yielding a 5\farcm2$\times$5\farcm2 field of view.  
Two adjacent fields are imaged in pairs (channels 1 and 3; channels 2 and 4) 
using dichroic beam splitters.  The telescope is then nodded to image a target
in all four channels.
\footnote{See Fazio et al.\ (2004) and the IRAC Users
Manual at http://ssc.spitzer.caltech.edu/irac/descrip.html}
We used exposure times of 30~s and a 5-position
medium-sized (52 pixels) dither pattern repeated one to four times. 
The total observing time per target ranged from 11 to 26~minutes.
The full array was read out.

The data were reduced using the post-basic-calibration data mosaics
generated by version 14 of the IRAC pipeline.\footnote{Further information 
can be found at http://ssc.spitzer.caltech.edu/irac/dh/}
The mosaics were flat-fielded and flux-calibrated using super-flats and global primary
and secondary standards observed by $Spitzer$.  We performed aperture photometry 
on the L and T dwarfs using apertures of radii 3--7 pixels.  The dwarfs were usually
isolated and well detected, so apertures of 7 pixels could be used.  We applied 
channel-dependent aperture corrections as described in Chapter~5 of the IRAC Data
Handbook.$^3$  For channels 1 to 3, the correction for the 7-pixel aperture is 3\%,
for the 5-pixel aperture it is 5--6\% and for the 3-pixel aperture it is 11--12\%. For
channel 4, these corrections are 4\%, 7\%, and 22\%.
We derived photometric errors from the uncertainty images that are provided
with the post-basic-calibration data.  The photometry was converted from milliJanskys to 
magnitudes on the Vega system using the zero-magnitude fluxes given in the IRAC Data
Handbook (280.9, 179.7, 115.0, and 64.1 Jy for channels 1 to 4, respectively).  
The fluxes, magnitudes, and errors for the observed dwarfs are given in Table~3.
Note that in addition to the errors in Table~3, there are absolute calibration uncertainties 
of 2--3\%.   There are also similarly-sized systematic uncertainties introduced
by pipeline dependencies, as we found by comparing the photometry produced by
pipeline versions 12, 13 and 14 for seven of our targets.
Photometry produced by versions 12 and 14 
agreed  well, to typically 4\%, however version 13 produced fainter magnitudes than
the other two pipelines, by $\sim$7\%.  The Spitzer helpdesk team recommend
version 14 of the pipeline, and suggest that version 13 used an over-aggressive
pixel rejection algorithm.
We adopt the total photometric uncertainty to be the sum in quadrature
of the values given in Table~3 plus 3\%.

IRAC flux densities follow the convention established by the {\it Infrared Astronomical Satellite 
(IRAS)} and other missions.  The IRAC flux density at a nominal wavelength assumes that the 
source has a spectrum described by $F_{\nu} \propto \nu^{-1}$.  However, this assumption is 
not valid for the SEDs of late-L and T dwarfs 
(Figure~1); $\lambda\times F_{\lambda}$ is not constant in these cases.   Hence the fluxes 
given in Table~3 are not representative of the nominal wavelengths given in Table~2.  For 
example, the 3.55~$\mu$m and 7.87~$\mu$m IRAC fluxes of the L9 dwarf shown in Figure~1 differ 
by $\sim 30$\% and $\sim 10$\%, respectively, from the actual fluxes at those wavelengths 
measured by {\it Spitzer's} IRS.  If the 
source spectrum is known, the quoted IRAC flux for a nominal wavelength $\lambda_0$ can be 
expressed as 
$$f_\nu^{IRAC}(\lambda_0) =
\int(\nu_0/\nu)f(\nu)S(\nu)\,d\nu \Bigg/ \int(\nu_0/\nu)^2S(\nu)\,d\nu $$
where $f(\nu)$ is the source spectrum and $S(\nu)$ is the system response function 
(Cushing et al.\ 2006).  

To put our data into context, we augment our sample with  
the IRAC photometry of M, L, and T dwarfs from Patten et al.\ (2006, 
hereafter P06), and include the photometry of two T dwarfs discovered in
IRAC images by Luhman et al. (2006).  

\section{Colors and Spectral Types}

\begin{figure} \includegraphics[angle=0,scale=.40]{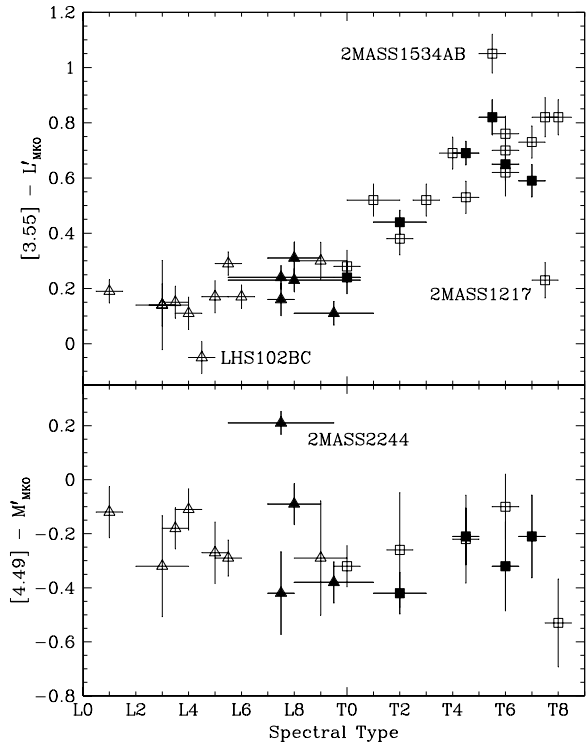} 
\caption{Comparison of MKO $L^{\prime}$ ($3.75~\mu$m) and $M^{\prime}$
($4.70~\mu$m) magnitudes with the IRAC channel 1 ([3.55]) and channel 2
([4.49]) magnitudes.  Triangles represent L dwarfs and squares represent T dwarfs.
Filled symbols are data presented in this work; open symbols are data
from Leggett et al.\ (2002), Reid \& Cruz (2002), G04, and P06.  
The following dwarfs are identified with abbreviated names: 
2MASS J12171110-0311131, 2MASS J1534498-295227AB and 2MASS J2244316+204343.}  
\end{figure}

Figure~2 shows the differences between IRAC channel 1 (3.55~$\mu$m) and
MKO $L'$ (3.75~$\mu$m) magnitudes and between IRAC channel 2 
(4.49~$\mu$m) and MKO $M'$ (4.70~$\mu$m) magnitudes, as functions of 
spectral type. The data 
are from this work, Leggett et al.\ (2002), Reid \& Cruz (2002), 
G04, and P06.  IRAC channels 1 and 2 span 
wavelengths similar to those of $L^{\prime}$ and $M^{\prime}$, respectively 
(Figure 1), and correlations can be used to check the consistency of 
{\it Spitzer} and ground-based data. 

Figure~2 shows that the IRAC channel~1 magnitudes (denoted [3.55]) become 
fainter than $L'$ magnitudes for later spectral types.  This behavior is expected 
because channel~1 samples more of the strong center of the 3.3~$\mu$m
CH$_4$ absorption band, and less of its long-wavelength wing 
and flux peak near 4~$\mu$m, compared to the $L^{\prime}$ bandpass (see Figure~1).  
The T7.5 dwarf 2MASS J12171110--0311131 is clearly anomalous in this plot; 
we discuss it further in \S4. The difference between IRAC channel~2 ([4.49]) and $M'$ is fairly 
constant ([4.49]--$M'$~$\approx -0.3$) for all types, with the exception
of the extremely 
red late-L dwarf 2MASS J22443167+2043433 (\S 4).  Both channel~2 and $M'$ 
sample H$_2$O and CO bands, even among T dwarfs (\S 4).
Although channel~2 is much wider than $M^{\prime}$, the filters are 
similarly sensitive to changes with spectral type.

\onecolumn

\begin{figure} \includegraphics[angle=0,scale=.7]{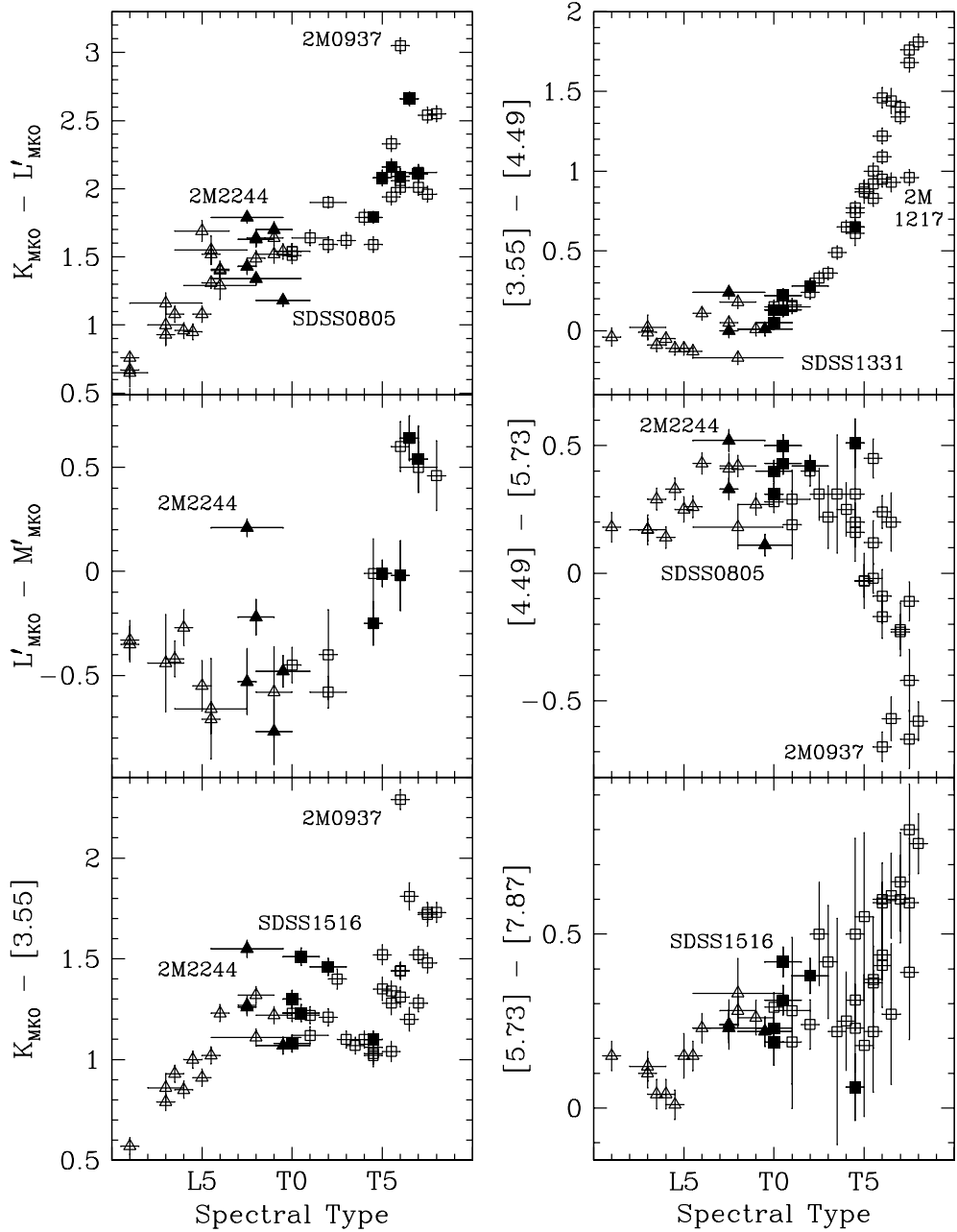} 
\caption{Colors as a function of spectral type.  The symbols are described in 
Figure 2.  Here, and in the following figures, open symbols are data from  Leggett et al.\ (2002), 
Reid \& Cruz (2002), 
G04, P06 and Luhman et al. (2006).
The following unusually blue or red dwarfs are identified and discussed 
in the text: 2MASS J0937347+293142, 2MASS J12171110-0311131,
2MASS J2244316+204343, SDSS J080531.80+481233, SDSS J133148.90-011651.4,
SDSS J151643.01+305344.4.}
\end{figure}

\twocolumn

Figure~3 shows various IRAC and MKO $KL'M'$ colors plotted as 
functions of spectral type.  The data are taken from this work, Leggett
et al.\ (2002), Reid \& Cruz (2002), K04, G04, C06, Luhman et al. (2006) and P06.  All of the
colors show scatter at the $\sim$0.5~mag level from mid-L to early-T types,
but general trends and some peculiarities can be seen.  2MASS J22443167+2043433, 
the reddest L dwarf currently known (Dahn et al.\ 2002; K04), is clearly 
anomalous in many colors.  SDSS J080531.80+481233.0 and SDSS J133148.90-011651.4, 
both unusually blue in the near-infrared (K04), are also anomalously blue in the 
IRAC  [4.49]--[5.73] and [3.55]--[4.49] colors.  SDSS J151643.01+305344.4, a red 
early-T dwarf (C06), is also red in $K$--[3.55] and [5.73]--[7.87].  Thus, the
unusual traits seen in these dwarfs at shorter wavelengths often
persist at longer wavelengths.  2MASS J09373487+2931409, which is extremely
red in $K$--$L'$ and $K$--[3.55], has suppressed $K$-band flux most likely due 
to very strong H$_2$ absorption (Burgasser et al.\ 2002a). 

\section{Interpretation of Colors and Luminosities}

In this section, we compare the observed IRAC and MKO colors with the predicted
colors from model atmospheres, and investigate the physical parameters that can 
produce the observed trends and variations seen in Figure~3.

\subsection{Description of Model Atmospheres}

State-of-the-art models of ultracool-dwarf atmospheres have been developed by members 
of our team (Ackerman \& Marley 2001; Marley et al.\ 2002; Saumon et al.\ 2003; 
and will be furthered described in an upcoming paper).
These models yield temperature-pressure-composition structures
under conditions of radiative-convective equilibrium using the thermal radiative transfer 
source function technique of Toon et al.\ (1989).  The gas opacity database includes the 
molecular lines of H$_2$O, CH$_4$, CO, NH$_3$, H$_2$S, PH$_3$, TiO, VO, CrH, FeH, CO$_2$, HCN, 
C$_2$H$_2$, C$_2$H$_4$, and C$_2$H$_6$, the atomic lines of the alkali metals (Li, Na, K, 
Rb and Cs), and continuum opacity from H$_2$ collisionally induced absorption (CIA), H$_2$,
H and He Rayleigh scattering, bound-free opacity from H and H$_2^+$, and free-free opacity 
from He, H$_2^-$, and H$_2^+$.  The models also allow opacity from arbitrary Mie-scattering 
particles.  

The radiative transfer code is closely coupled to cloud models that use a sedimentation 
efficiency parameter, $f_{\rm sed}$, to describe the balance between the downward 
sedimentation of condensates and the upward convection that replenishes the grains.
Large values of $f_{\rm sed}$ correspond to rapid growth of large grains, which then 
quickly fall out of the atmosphere and yield physically and optically thin clouds.  
When $f_{\rm sed}$ is small, the grains grow more slowly and the clouds are thicker.
The models have been successfully applied to the atmospheres of brown dwarfs and 
Jupiter (Ackerman \& Marley 2001; K04; G04).  

The models have been further advanced to include vertical
mixing, which draws up chemical species from the well-mixed, deep, hot layers into
the cool radiative photosphere.  Although the photosphere would be expected to
be stable and in chemical equilibrium, even a small degree of mixing (such
as is commonly observed in the stratospheres of the solar system's giant planets) enhances the
abundances of stable species like
CO and N$_2$  relative to their chemical equilibrium values, 
while those of H$_2$O, CH$_4$ and NH$_3$
are reduced (Fegley \& Lodders 1996, Lodders \& Fegley 2002, Saumon et al.\ 2003, 2006).
This phenomenon has been observed in planetary atmospheres
(Noll et al.\ 1988, Noll \& Larson 1991, Fegley \& Lodders 1994, B\'ezard et al.\ 2002)
and in brown dwarfs (Noll et al.\ 1997,
Griffith \& Yelle 1999, Saumon et al.\ 2000, 2006). In the convection zone that forms in
the lower atmosphere, mixing is vigorous and occurs on a time scale $\tau_{\rm mix}=H_p/v_{\rm conv}$
where $H_p$ is the pressure scale height and $v_{\rm conv}$ is the convective velocity, obtained from
the mixing length theory.  The physical process of mixing in the radiative zone is not known but
it should take place on a much longer time scale.  For lack of a theory of mixing in brown dwarf
atmospheres, it is assumed to be a diffusive process and the mixing time scale is given by
$\tau_{\rm mix}=H^2_p/K_{zz}$, where $K_{zz}$ is the coefficient of eddy diffusivity,
considered a free parameter here.  For the present study, models
have been calculated with $K_{zz}=0$ (no mixing, chemical equilibrium),
$10^2$ and $10^4$~cm$^2$~s$^{-1}$ in the radiative zone.  For comparison, the mixing
time scale in the convection zone would correspond to values of $K_{zz} \approx 10^7$--$10^9$~cm$^2$~s$^{-1}$.

We have generated a large grid of models with ranges of $T_{\rm eff}$, surface gravity 
$g$, $f_{\rm sed}$, metallicity [M/H], and mixing coefficient $K_{zz}$.  
In this paper we treat the dynamical parameters,  $K_{zz}$ and $f_{\rm sed}$, as independent
variables, in the same way as we treat $g$ and $T_{\mathrm eff}$.
However,  the global properties of each dwarf must ultimately define the
true properties of the cloud, and the vigor of atmospheric mixing.  
Two otherwise identical L dwarfs, for example, could
not exhibit two different sedimentation efficiencies and eddy diffusion coefficients.  
Eventually we will be able to treat $K_{zz}$ and $f_{\rm sed}$ as
functions of other physical properties, but this cannot happen until we 
understand how differences in metallicity, rotation rate, effective temperature, and
gravity impact sedimentation and mixing in the atmosphere. Thus for the time being we treat
$K_{zz}$ and $f_{\rm sed}$ as independent variables, and use them  to understand the ranges of 
behaviour manifested by the known objects.

Saumon et al.\ (2006) have validated the
models by reproducing the entire red through mid-infrared spectrum of the T7.5 dwarf
Gl~570D. Cushing et al.\ (in preparation) also find very good agreement between
the model spectra and the observed $zJHKL^\prime$ and IRS spectra of L0--T4 
dwarfs.  Work in progress comparing the calculated and 
observed near-infrared colors of L and T dwarfs shows that L dwarfs 
are well represented by models with $f_{\rm sed}=1$--2, while T dwarf colors are well 
matched by models with $f_{\rm sed}=4$ or by cloudless models (i.e., the clouds lie 
below the photosphere at all wavelengths).  Finally we note that  because the 
models are complex and CPU-intensive, not all combinations of all parameters are 
available at this time.

\subsection{Comparison of Models and Data}

Figures 4, 5, and 6 each show four color--color diagrams that compare new and previously
published 2.2--7.9~$\mu$m photometry with synthetic colors generated from 
models covering appropriate ranges of $T_{\rm eff}$, $f_{\rm sed}$, log~$g$, 
and $K_{zz}$.  Generally, the model sequences are generated for 
$500 \leq T_{\rm eff}\leq 2400$~K in steps of 100~K.  However, the 
dustiest ($f_{\rm sed}=1$ and 2) models with vertical mixing ($K_{zz}=10^2$ 
and $10^4\,$cm$^2$~s$^{-1}$) are generated for the smaller range 
$800 \leq T_{\rm eff}\leq 2400$~K because calculating cold, dusty, atmospheres 
with mixing is difficult and CPU intensive.  This limitation does not affect our
analysis or conclusions because the condensate cloud decks are expected to 
lie below the photosphere for $T_{\rm eff}\lesssim 1000\,$K.  Consequently,
we do not consider inappropriate dusty models for mid- to late-T dwarfs or 
cloudless models for early- to mid-L dwarfs.  All model sequences assume solar 
metallicity, except for one metal-poor cloudless sequence.

Figure 4 demonstrates the colors' sensitivity to gravity for 
the two extremes of cloud properties: a model with 
$f_{\rm sed}=1$ and a completely cloudless model.  Both models assume no vertical 
mixing.  The gravities 
shown ($\log g=4.5$, 5.0 and 5.5) correspond to $\sim 15$--75 Jupiter 
masses for ages greater than 0.3~Gyr (Burrows et al. 1997), which are 
appropriate for our field sample with 700 $\lesssim T_{\rm eff}\lesssim 2200\,$K.
Figure~5 shows sequences for $\log g=5.0$ and a full range of sedimention parameters, 
$f_{\rm sed}=$1, 2, 3, 4, and cloudless, without vertical mixing.  It also shows a 
metal-poor ([M/H]$=-0.3$), $\log g=5.0$, cloudless sequence.  Figure~6 shows 
$f_{\rm sed}=$1, 2, 3, and cloudless sequences for $\log g=5.0$, with vertical 
mixing parameters of $K_{zz}=10^2$ and $10^4\,$cm$^2$~s$^{-1}$.

In general, the varying colors of brown-dwarf model spectra are not straightforward 
functions of $T_{\rm eff}$, gravity, composition, and $f_{\rm sed}$.  One must also
consider the variations in chemistry and overall gas and cloud opacity that affect 
the depth of the ``photosphere.''  The relative strengths of absorption bands in the 
spectrum are complex functions of all these parameters.  For the sake of brevity and
clarity, we will explain the trends for those cases where the interpretation is clear.

\subsubsection{Gravity}
Figure~4 indicates that gravity only impacts significantly (at the $\wig>$20\% level)
the $K$--$L'$, $K$--[3.55], and [4.49]--[5.73]
colors of mid- to late-T dwarfs.  The [5.73]--[7.87] color is not 
significantly impacted, nor are any of the colors of dusty or cloudless dwarfs earlier than
type T6 (although the absolute magnitudes can differ, as we discuss in \S 5).

\onecolumn

\begin{figure} \includegraphics[angle=0,scale=0.7]{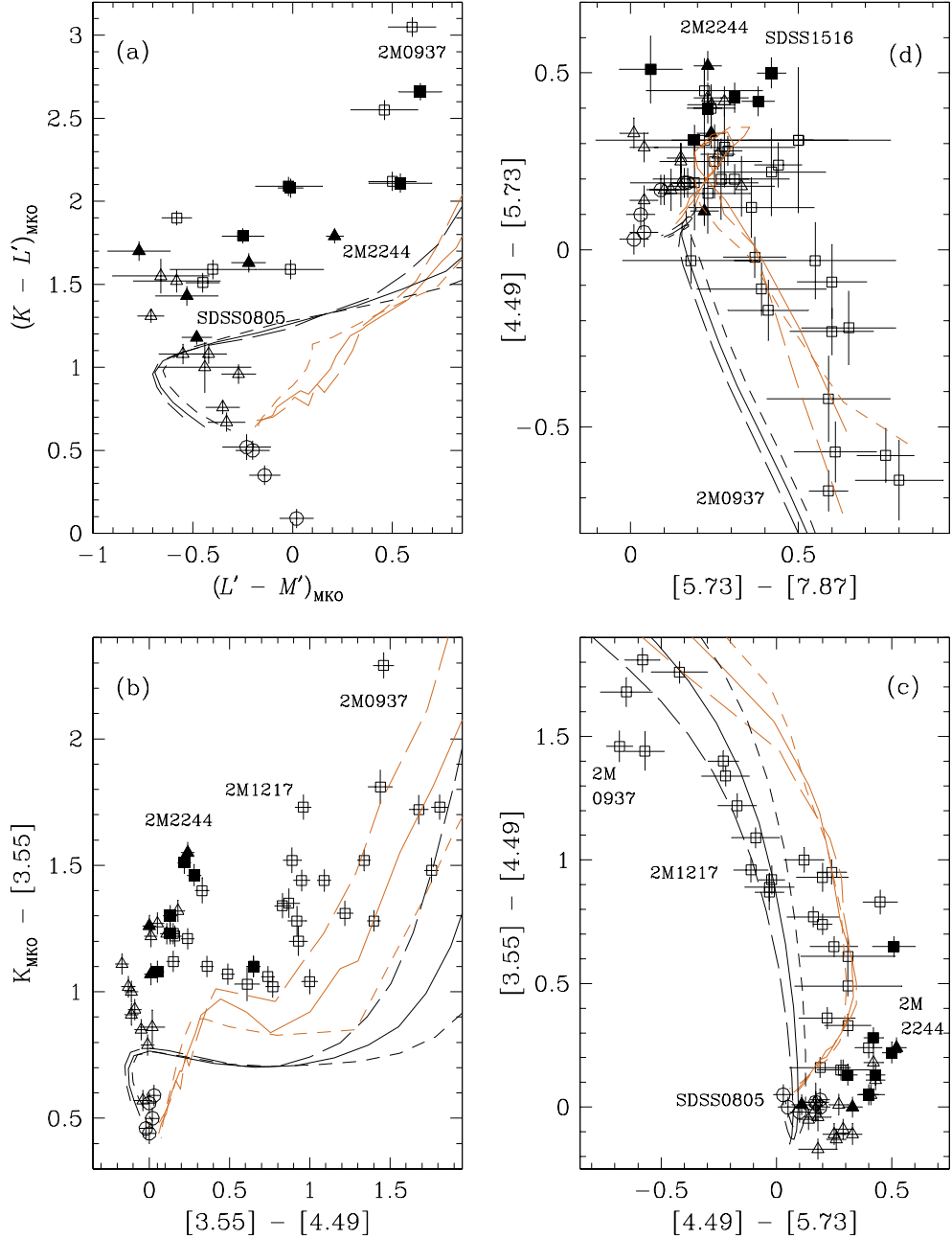}
\caption{\footnotesize{Color--color diagrams and the effect of gravity: (a) $K$--$L^{\prime}$ vs.\ 
$L^{\prime}$--$M^{\prime}$, (b) $K$--[3.55] vs.\ [3.55]--[4.49], (c) [3.55]--[4.49] vs.\ 
[4.49]--[5.73], and (d) [4.49]--[5.73] vs.\ [5.73]--[7.87].  Circles represent M dwarfs, 
triangles represent L dwarfs, and squares represent T dwarfs.  These dwarfs are identified: 
2MASS J0937347+293142, 2MASS J12171110-0311131, 2MASS J2244316+204343, SDSS J080531.80+481233
and SDSS J151643.01+305344.4.  The curves show the synthetic colors as a function of $T_{\rm eff}$
for two cloud models ($f_{\rm sed}=1$ -- orange curves; cloudless -- black curves), three gravities
($\log g = 4.5$ -- short-dashed curves; $\log g = 5.0$ -- solid curves; $\log g = 5.5$ -- 
long-dashed curves), and fixed solar metallicity.
In panels (a)--(c), $T_{\rm eff} = 2400$~K lies at the bottom and 
$T_{\rm eff}$ decreases upward.  In panel (d) the color sequence progresses from left to
right as  $T_{\rm eff}$ decreases.}}
\end{figure}

\twocolumn

For $T_{\rm eff} \wig< 1000\,$K, the
$K$-band flux is sensitive to gravity because it corresponds to a peak in the H$_2$ CIA
opacity.  This opacity depends quadratically on the gas number density, unlike most other 
opacity sources that are roughly linear functions of the density.  Higher gravities cause
higher photospheric pressures and a greater CIA opacity, which depresses the $K$-band flux.
Figures 4a and 4b indeed show redder $K$--$L^\prime$ and $K$--[3.55] colors as 
$\log g$ increases.  Only the coolest T dwarfs are affected, however, because at 
higher $T_{\rm eff}$ the H$_2$O opacity in the $K$-band increases and CIA is no longer 
dominant.  

The effect of gravity on the IRAC colors (Figures 4c and 4d) is more subtle, and the
cause varies with $T_{\rm eff}$ or spectral type.  The chemistry of CH$_4$ and PH$_3$, and the 
temperature-pressure profile, are all dependent on gravity. The net result is 
bluer [4.49]--[5.73] colors as gravity increases, for $T_{\rm eff} <$1400~K, or late-L and 
all T types.

\subsubsection{Sedimentation efficiency}

Figure~5 shows that the sedimentation efficiency, which regulates the thickness of 
the cloud decks and the geometric depth of the photosphere, has a large effect on the
near- and mid-infrared colors of ultracool dwarfs.  The wavelength dependence of the 
cloud opacity is fairly flat in the infrared (Ackerman \& Marley 2001), except for a 
possible silicate feature around $10\,\mu$m (Cushing et al. 2006).  Hence, the cloud 
affects all bands from the 
optical to the mid-infrared.  At longer wavelengths, the gas opacity tends to dominate 
and the mid-infrared flux is less sensitive to the choice of $f_{\rm sed}$, except when 
the cloud is very thick ($f_{\rm sed} \approx 1$).

The range of $f_{\rm sed}$ depicted in Figure~5 lies between the cloudless and thick-cloud 
extremes shown in Figure~4, and so the model sequences have intermediate colors.  All the 
model color sequences in Figure~5 show a sharp reversal at $T_{\rm eff} \approx 1300$~K.
This reversal is also seen in the observed colors at the L--T transition 
(triangle to square symbols in the plots), where the atmospheres
are changing rapidly from cloudy to cloudless.  The 
thick-cloud model ($f_{\rm sed}=1$) generally gives the closest fit to all the color-color
diagrams, except for $K$--$L^\prime$ vs.\ $L^\prime$--$M^\prime$, which is better fit with
cloudless models.  Nonetheless, a significant mismatch remains, which compels us to explore 
model parameters other than gravity and $f_{\rm sed}$, 
in search of better agreement with the observations.

\subsubsection{Metallicity}

Figure~5 shows that decreasing the metallicity of the cloudless models by 0.3 dex
impacts only the colors of the latest T dwarfs ($T_{\rm eff} \wig< 1000\,$K) between 
2.2 and 3.8~$\mu$m.  Reducing the metallicity has a qualitatively similar effect on 
the chemistry of C, N, and O as increasing the gravity or gas density (Lodders \& Fegley 2002).  
The abundances of molecules composed of two metals (e.g. CO) decrease much faster than 
the abundances of molecules with only one metal (e.g. H$_2$O).  Thus, CO and N$_2$ will be 
relatively less abundant than H$_2$O, CH$_4$ and NH$_3$.  In addition, a decrease in the
metallicity reduces the H$_2$O abundance and the overall opacity in the atmosphere
(which is dominated by H$_2$O absorption), and results in increased gas density.
Finally, the importance of H$_2$ CIA increases in a metal poor atmosphere.
All these effects mimic the effect of increasing gravity.  The [M/H]$=-0.3$, $\log g=5$ 
sequence (Figure~5) is very similar to the [M/H]=0, $\log g=5.5$ sequence (Figure~4)
in all the color-color diagrams shown.  In a future paper, we will explore whether 
decreasing metallicity or increasing gravity in dusty L dwarfs has effects similar 
to those of increasing sedimentation efficiency (i.e., thinning of the cloud decks),
and vice versa.

\subsubsection{Vertical mixing}

Figures 4 and 5 show that the models with reasonable ranges of $\log g$, $f_{\rm sed}$, and 
[M/H] do not reproduce the observed colors of L and T dwarfs very well, although the
general trends with $T_{\rm eff}$ are satisfactory.  The worst mismatches are seen in the 
$K$--$L^\prime$ and/or $L^\prime$--$M^\prime$ colors (Figures 4a and 5a), where discrepancies
of up to a full magnitude are seen for the T dwarfs.  The synthetic [5.73]--[7.87] colors
(Figures 4d and 5d) are also $\sim 0.3$~mag too blue.  This discrepancy is consistent with 
the detailed fit of the SED of the T7.5 dwarf Gl~570D with the same models (Saumon et 
al.\ 2006) where the fluxes in the deep CH$_4$ and H$_2$O bands from 5.5 to 8~$\mu$m 
(Figure 1) are underestimated.  

\onecolumn

\begin{figure} \includegraphics[angle=0,scale=0.7]{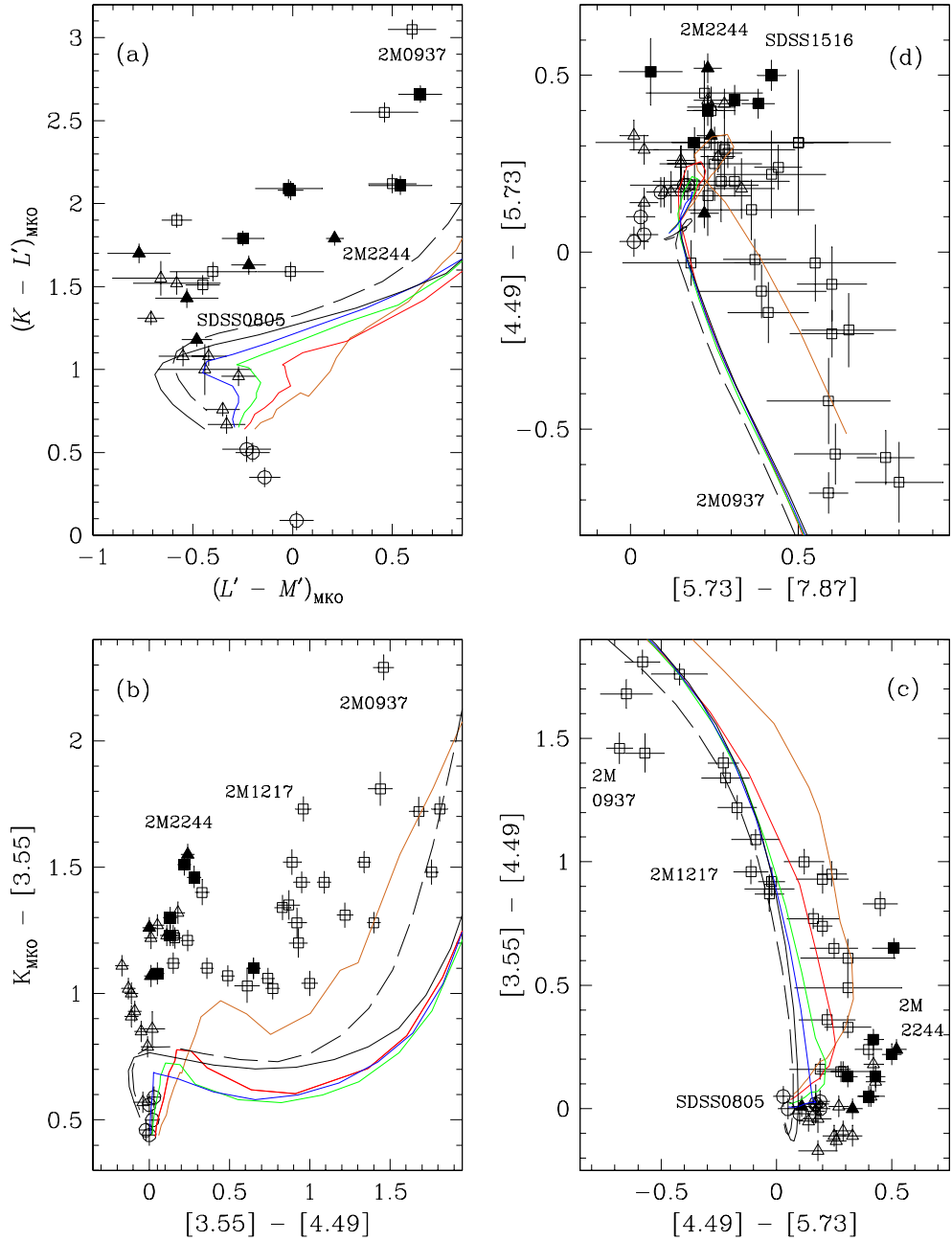}
\caption{Color--color diagrams and the effect of sedimentation efficiency.  The diagrams and symbols
are as in Figure~4.  The curves show the synthetic colors as a function of $T_{\rm eff}$
for five cloud models ($f_{\rm sed}=4$ -- blue curves; $f_{\rm sed}=3$ -- green curves; 
$f_{\rm sed}=2$ -- red curves; $f_{\rm sed}=1$ -- orange curves; cloudless -- black curves), two metallicities
(solar -- all solid curves; [M/H]$=-0.3$ -- dashed curve), and fixed $\log g = 5.0$.}
\end{figure}

\twocolumn

It is also puzzling that the wide ranges of parameters considered so far are
unable to reproduce the range in the observed colors of L dwarfs, shown as triangles
in Figures 4 and 5. For example, the observed range in $K - L^{\prime}$ is 
$\sim$~1 magnitude, and in [4.49] -- [5.73] it is $\sim$~0.5 magnitudes (Figure 3). The
synthetic colors for the appropriate range of $T_{\rm eff}$ (1400~K -- 2400~K, G04)
only vary by $\sim$~0.6 magnitudes at $K - L^{\prime}$ and  by  $\sim$~0.3 magnitudes at
[4.49] -- [5.73], for $4.5 \leq \log g \leq 5.5$ and $1 \leq f_{\rm sed} \leq 4$.

As described in \S 4.1,
there is growing evidence that non-equilibrium chemistry caused by vertical mixing is common 
in T dwarfs.  The spectrum of the T6 dwarf Gl 229B shows enhanced CO and depleted NH$_3$ 
(Noll et al.\ 1997; Oppenheimer et al.\ 1998; Griffith \& Yelle 1999; Saumon et al.\ 2000),
and the mid-infrared spectrum of the T7.5 dwarf Gl 570D also shows depleted NH$_3$ 
(Saumon et al.\ 2006).  Models incorporating non-equilibrium chemistry have been shown to
reproduce the $M'$ photometry of brown dwarfs better than equilibrium models (Leggett et al.\
2002; G04).  Another example is shown in Figure 1: the 3$\,\mu$m CH$_4$ band in the spectrum of 
the L9 dwarf 2MASS J09083803+5032088 is much weaker than predicted by chemical equilibrium 
models with parameters appropriate for late-L field dwarfs ($T_{\rm eff}=1400$~K, 
$f_{\rm sed}=2$, and $\log g=5$). 

Figure~6 shows non-equilibrium models computed with two mixing efficiencies, $K_{zz}=10^2$ 
and $10^4\,$cm$^2$~s$^{-1}$.   The synthetic colors are very sensitive to vertical mixing,
because the filter bandpasses sample absorption features caused by molecules whose abundances
are significantly affected by the mixing.  Figure~1 shows that $K$, $L^{\prime}$, and IRAC 
channels 1 and 4 sample CH$_4$ bands; $K$, $M^{\prime}$, and IRAC channel 2 sample CO bands;
and all filters (especially IRAC channels 2 and 3) are affected by H$_2$O bands (Saumon et 
al.\ 2003).  The red part of the IRAC channel 4 flux is also affected by NH$_3$ absorption 
in the T dwarfs.  Non-equilibrium effects on the CO and CH$_4$ bands appear below $T_{\rm eff}
\sim 1600$~K (or spectral types later than L5), at which point carbon is no longer fully 
sequestrated in CO.  All four diagrams in Figure~6 show that the 2--8~$\mu$m photometry of 
late-L and T dwarfs is well reproduced by models that include vertical mixing, especially 
when $T_{\rm eff} \wig < 1400\,$K (or spectral types later than about L7).  Most of the 
data are matched by models with $K_{zz} \sim 10^4$~cm$^2$~s$^{-1}$, with residuals of 
$\lesssim 20$\%.  Furthermore, the observed range in color for L and T types can 
be accommodated by variations in $K_{zz}$.

The best match to the $K$--$L^\prime$ vs.\ $L^\prime$--$M^\prime$ diagram (Figure~6a) is
obtained with the model parameters $K_{zz}=10^4\,$cm$^2$~s$^{-1}$, $\log g=5$, [M/H]=0, 
and $f_{\rm sed}\ge 2$ (for $T_{\rm eff} \gtrsim 1400$~K) or no clouds (for $T_{\rm eff}
\lesssim 1400$~K).  The same parameters apply to the $K$--[3.55] vs.\ [3.55]--[4.49] 
diagram (Figure~6b), except that $f_{\rm sed}=2$ provides the best fit over the entire 
range of $T_{\rm eff}$.  Higher values of $f_{\rm sed}$ give equally good agreement 
if the gravity is increased.  The models are less successful in reproducing the 
[3.55]--[4.49] vs.\ [4.49]--[5.73] diagram (Figure~6c).  Vertical mixing has little effect
in this diagram, and $f_{\rm sed} \approx 1$--2 best matches the colors of the M and L 
dwarfs.  Further improvement at the warm end of the color sequence can be achieved with 
above-solar metallicity at the expense of a worse fit at T-dwarf temperatures.
Finally, the [4.49]--[5.73] vs.\ [5.73]--[7.87] diagram (Figure~6d) shows 
that the  observed colors of the hotter objects 
([4.49]--$[5.73] \wig> 0.1$) are best reproduced with $f_{\rm sed}=1$ and 
$K_{zz}=10^4\,$cm$^2$~s$^{-1}$.  

To summarize, good agreement with all four color-color diagrams is obtained with models with 
solar metallicity, $\log g \approx 5$, $f_{\rm sed}=$1--2 for $T_{\rm eff} > 1400\,$K, and 
cloudless models at lower $T_{\rm eff}$, and an eddy diffusion coefficient of 
$K_{zz}=10^4\,$cm$^2$~s$^{-1}$.  The data strongly support the idea that non-equilibrium 
chemistry, as modelled by vertical transport, is prevalent in late-L and T dwarfs.

\subsubsection{Objects with unusual colors}

Several objects stand out in Figure 3 by being unusually blue or red 
for their spectral type in one or more colors.  We use the model
sequences in Figures 4--6 to make a preliminary assessment of these
peculiar colors.

\begin{itemize}

\item

The L7.5 dwarf 2MASS J22443167+2043433 is very red in $K$--$L^\prime$, $L^\prime$--$M^\prime$,
$K$--[3.55], and [4.49]--[5.73], suggesting that its atmosphere is very cloudy 
($f_{\rm sed}<1$) and strongly affected by vertical mixing ($K_{zz}>10^4\,$cm$^2$~s$^{-1}$),
as shown for example by Figure 6(b).

\end{itemize}

\onecolumn

\begin{figure} \includegraphics[angle=0,scale=0.7]{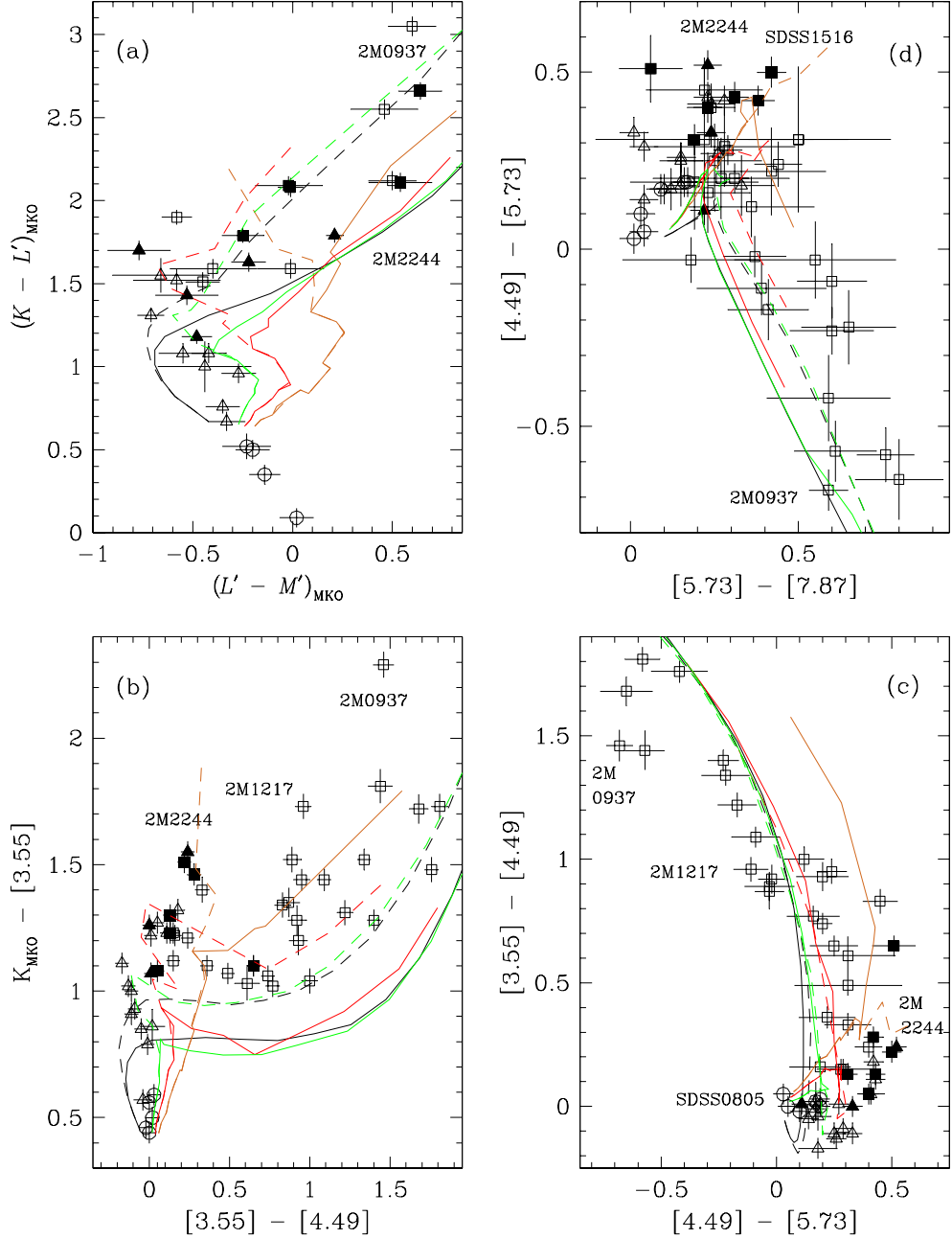}
\caption{Color--color diagrams and the effect of vertical mixing.  
The diagrams and symbols are as in Figure~4.  
The curves show the synthetic colors as a function of $T_{\rm eff}$
for four cloud models ($f_{\rm sed}=3$ -- green curves; $f_{\rm sed}=2$ -- red curves; 
$f_{\rm sed}=1$ -- orange curves; cloudless -- black curves), two eddy diffusion
coefficients ($K_{\rm zz}=10^2$~cm$^2$~s$^{-1}$ -- solid curves; $K_{\rm zz}=10^4$~cm$^2$~s$^{-1}$ 
-- dashed curves), solar metallicity, and $\log g = 5.0$.}
\end{figure}

\twocolumn

\begin{itemize}
\item

SDSS J133148.90-011651.4 is an L8 dwarf that is blue in [3.55]--[4.49] and [4.49]--[5.73], 
suggesting that it has a high value of $K_{zz}$ and thin clouds with $f_{\rm sed}\wig>3$
(Figure 6(b)).  

\item

SDSS J080531.80+481233 (L9.5) is blue in [4.49]--[5.73]
and $K$--$L^\prime$, and is similar to SDSS J133148.90-011651.4.  It is
best reproduced by a high sedimentation model,
with $f_{\rm sed}\sim 4$ or perhaps no clouds, and modest vertical mixing.

\item

The T0.5 dwarf SDSS J151643.01+305344.4
is red in $K$--[3.55] and  [5.73]--[7.87],
which requires both vertical mixing and fairly thick clouds with $f_{\rm sed} \le 2$,
as shown for example by Figure 6(d).

\item

The well-known peculiar T6 dwarf 2MASS J09373487+2931409 has such extremely red  
$K$--$L^\prime$ and $K$--[3.55] colors that it can only be explained with a 
high gravity {\it and} low metallicity atmosphere, plus vertical mixing (e.g.
Figures 4(b), 5(b) and 6(b)).

\item

2MASS J12171110-0311131 (T7.5) is very blue in [3.55]--[4.49], moderately red
in [4.49]--[5.73], and moderately blue in other colors (P06).
This dwarf has a bright $K$-band flux peak, implying weak H$_2$ 
CIA and therefore low gravity (Burgasser et al.\ 2006a).
However, the color sequences in Figures 4 and 5 show that decreasing gravity and 
increasing metallicity would not reproduce the observed colors, so the high $K$ 
flux may be due to something other than low H$_2$ opacity.  The colors may be due 
to extremely efficient vertical mixing, e.g., $K_{zz}\sim 10^6\,$cm$^2$~s$^{-1}$,
which leads to reduced CH$_4$ and brighter $K$ and [3.55] magnitudes (see Figure 6b).  
This T dwarf has been imaged by $HST$ and appears to be single (Burgasser et al.\ 2006c),
so duplicity does not seem to explain its unusual colors.

\end{itemize}

As we note in \S 4.1, the above discussion implies that all five model parameters, 
$T_{\rm eff}$, gravity, [M/H], $f_{\rm sed}$ and $K_{zz}$, can
be individually varied to fit the properties of each unusual object.   In fact the dynamical
parameters $f_{\rm sed}$ and $K_{zz}$
will depend on the other three bulk properties, as well as on rotation and the detailed
chemical composition.  Further modeling of individual objects, such as
2MASS J22443167+2043433 and SDSS J133148.90-011651.4
which seemingly have similar spectral types and
vertical mixing but very different $f_{\rm sed}$, will help to illuminate such issues.

\subsection{Effects of Gravity and Vertical Mixing on Luminosity}

P06 suggest that the spread in the IRAC photometry of T dwarfs is due to variations
in gravity within the sample.  Except when they are very young, brown dwarfs cool at 
nearly constant radii and their gravity strongly correlates with their mass and their 
luminosity.  The effect of gravity on models is therefore better revealed in color--magnitude
diagrams than in color--color diagrams.  

Figure~7 shows the IRAC channel 1 and 2 absolute 
magnitudes as functions of color, $M_{[3.55]}$ vs.\ [3.55]--[4.49] and
$M_{[4.49]}$ vs.\ [4.49]--[5.73], for those objects which have measured parallaxes.
The data reveal clear sequences as a function of spectral type, with
increased scatter in the T dwarf regime.  Evolutionary calculations indicate that
for a population of very-low-mass stars and brown dwarfs older than $\wig> 1\,$Gyr 
(such as our present sample), objects with lower $T_{\rm eff}$ will span a wider range of 
mass and gravity (see Figure~1 of Saumon et al.\ 2006).    However, the model sequences
shown in Figure~7 indicate that, for the plausible range of $\log g=4.5$---5.5,
the width due to gravity of the modelled sequence is narrower than the observed T dwarf sequence,
at least for the ($M_{[4.49]}$) vs.\ [4.49]--[5.73] diagram.
We conclude that a range of gravities is not  sufficient  by itself
to explain the width of the T dwarf sequence in these diagrams.
Increasing the vertical mixing coefficient from $K_{zz}=0$  to
$10^4\,$cm$^2$~s$^{-1}$ shifts the T dwarfs' [3.55]--[4.49] colors blueward and the
[4.5]--[5.8] colors redward, by several tenths of a magnitude --- and
mostly recovers the observed sequence.
This large shift clearly shows that a range of vertical mixing efficiency,
($K_{zz} \sim 10^2$--$10^6\,$cm$^2$~s$^{-1}$) in cloudless brown-dwarf atmospheres 
could reproduce the observed range of T dwarf 3--8~$\mu$m colors.

\onecolumn
\begin{figure} \includegraphics[angle=-90,scale=.6]{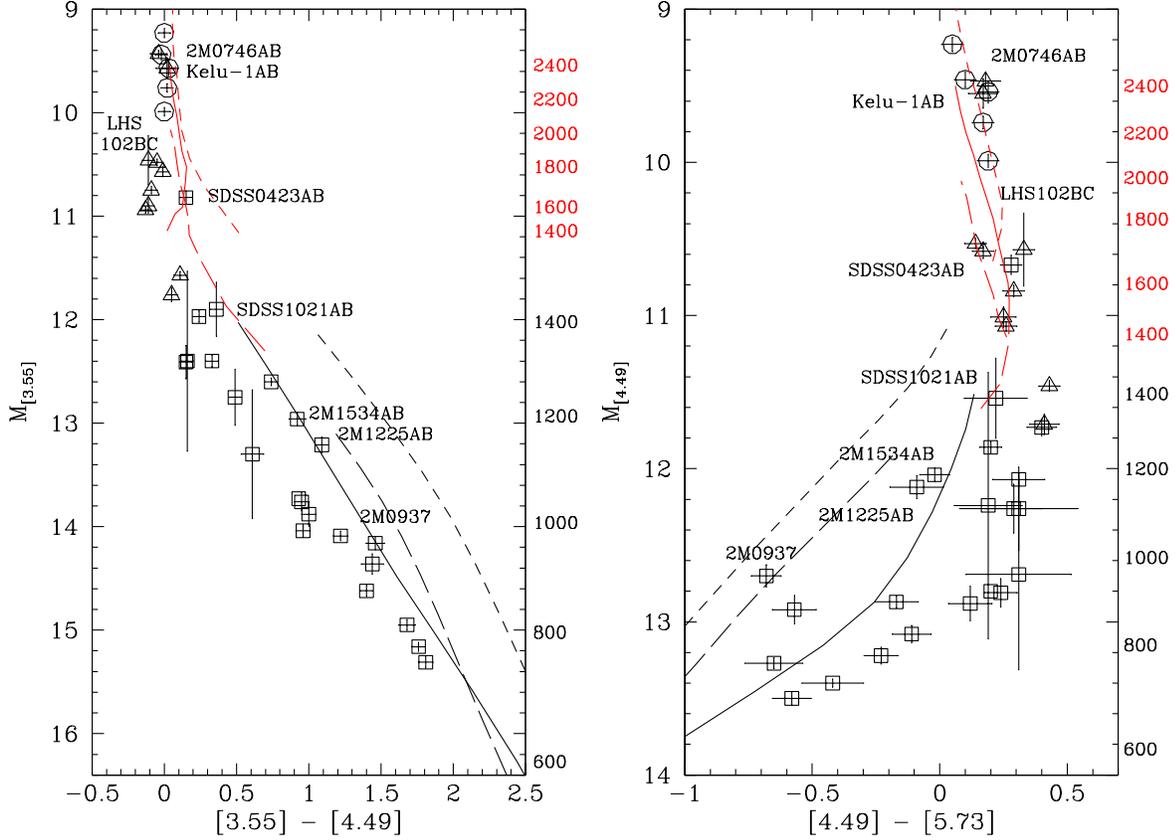}
\caption{IRAC color-magnitude diagrams for those objects with measured parallax.  
The symbols are described in Figure 4. Known binaries (with combined photometry) are
labelled, as is the peculiar T6 dwarf 2MASS J0937347+293142.   Cloudy models with 
$f_{\rm sed}=2$ and $1400 \leq T_{\rm eff} \leq 2400$~K are shown by the red curves, 
and cloudless models with $500 \leq T_{\rm eff} \leq 1400$~K
are shown by the black curves.  The plausible range of gravity for field brown dwarfs 
($\log g=4.5$--5.5) is bracketed by the short-dashed and long-dashed curves, respectively.  
These sequences do not include vertical mixing.  Models including vertical mixing, 
with $K_{zz}=10^4\,$cm$^2$~s$^{-1}$ and $\log g=5$, are shown by the solid curves. 
Labels on the right axes indicate $T_{\rm eff}$ (K) for the non-equilibrium models.}
\end{figure}
\twocolumn

\begin{figure} \includegraphics[angle=0,scale=0.4]{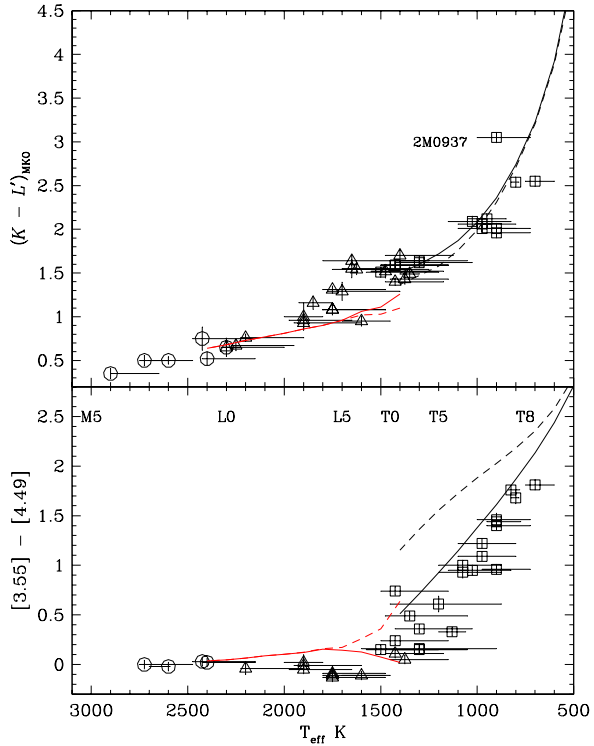} 
\caption{Colors as indicators of $T_{\rm eff}$.  The values of $T_{\rm eff}$
are determined from bolometric luminosities by G04 and Luhman et al. (2006); the G04 
values have been revised for 
the binarity of Kelu-1 (Liu \& Leggett 2005), LHS 102BC (Golimowski et al.\ 2004b), SDSS 
J042348.57-041403.5 (Burgasser et al.\ 2005) and SDSS J102109.69-030420.1 (Burgasser et al.\ 2006c).
The symbols are described in Figure 4.  Approximate spectral types are indicated, using the 
$T_{\rm eff}$ vs.\ spectral type relationship of G04.  The errors in $T_{\rm eff}$ are dominated by 
uncertainty in age.  The curves show model color sequences with [M/H] = 0, $\log g=5$, and 
cloudless atmospheres with $500 \leq T_{\rm eff} \leq 1400$~K (black curves) or cloudy
atmospheres with $f_{\rm sed}=2$ and $1400 \leq T_{\rm eff} \leq 2400$~K (red curves).  
The dashed curves are chemical equilibrium models ($K_{zz}=0$) and the solid curves are 
non-equilibrium  models with $K_{zz}=10^4$~cm$^2$~s$^{-1}$.  The peculiar T6 dwarf 
2MASS J0937347+293142 is identified. }
\end{figure}

\section{Effective Temperature Indicators}

Of the colors considered in this paper, two show promise as effective temperature
indicators.  Figure 3 shows that both $K$--$L^{\prime}$ and [3.55]--[4.49] 
exhibit relatively little scatter and vary monotonically with spectral type.  
The monotonicity of $K$--$L^{\prime}$ with spectral type was first noted by Stephens et 
al.\ (2001).  This relation was subsequently used by Chauvin et al.\ (2004) and 
Neuh{\"a}user et al.\ (2005) to assign a spectral type and $T_{\rm eff}$ to candidate 
planetary-mass companions of more massive objects.  

Figure~8 shows the observed 
$K$--$L^{\prime}$ and [3.55]--[4.49] colors as a function of $T_{\rm eff}$, as determined 
by G04 and Luhman et al. (2006).  
Cloudy ($f_{\rm sed}=$2) and cloudless model color sequences are also shown.  
The observed colors are well reproduced by the models in the appropriate range of 
$T_{\rm eff}$ when vertical mixing is included.   Although [3.55]--[4.49] is constant 
down to $\sim 1700$~K, the onset of CH$_4$ absorption moves the colors steadily redward 
through the late-L and the T sequences.  As shown in the figure, the putative Y 
dwarfs (i.e., dwarfs cooler than T dwarfs) will also likely be red in both colors.

We note that care must be taken when applying  empirical color--$T_{\rm eff}$
relations appropriate for field objects to young low-mass and low-gravity, and possibly
non-solar metallicity, objects.
For late-T dwarfs, H$_2$ CIA opacity is important, particularly in the $K$ band,
and extremely sensitive to gravity and metallicity (\S 4.2.1, 4.2.3).
For example, for a mid-T dwarf with $K$--$L^{\prime} \approx 2.5$, varying $\log g$ by 0.5 dex 
produces a change in $T_{\rm eff}$ of 100~K, and varying [M/H] by 0.3 dex a change of $\sim$50~K.
In addition, $K$--$L^{\prime}$ is sensitive to the mixing parameter $K_{zz}$ 
for late-L to early-T types, and [3.55]--[4.49] is also sensitive to this parameter 
for late-L to late-T types.  Figure 8 shows that if $K_{zz}$ is substantially larger or smaller
in the atmospheres of low-mass objects,
errors in derived $T_{\rm eff}$ of 100---300~K will result.

\section{Conclusions}

Our NIRI $L^{\prime}$ (3.75~$\mu$m), $M^{\prime}$ (4.70~$\mu$m), and IRAC 
channels 1--4 (3.55, 4.49, 5.73 and 7.87$\,\mu$m) photometry, when combined with 
previously published data, constitutes a sample of 30--50 objects (depending on 
the colors) that is large enough to establish the behaviour of the entire L--T 
spectral sequence in color--color diagrams.
Useful comparisons with synthetic mid-infrared colors are now possible. Current 
model atmospheres are quite complex, incorporating detailed molecular
and atomic opacities, the formation and sedimentation of condensates, and
vertical mixing leading to non-equilibrium chemistry. 

Generally the models reproduce the observed mid-infrared colors well.  For
early- to mid-L dwarfs ($T_{\rm eff} \gtrsim 1600$~K), the synthetic IRAC colors
are insensitive to variations in the model parameters, except 
for the case of thick clouds ($f_{\rm sed} = 1$).
The colors of the late-L and T dwarfs are only reproduced by
models that include non-equilibrium chemistry arising from
vertical transport (mixing) in the radiative region of the atmosphere.  
Efficient mixing is necessary, with an eddy diffusion coefficient of 
$K_{zz} \approx 10^4\,$cm$^2$~s$^{-1}$.
The comparisons presented here show that vertical mixing in the atmospheres of late-L and T 
dwarfs ($T_{\rm eff} \wig< 1600\,$K) affects the chemistry of carbon and is 
widespread among these objects.

Late-L and early-T dwarfs that are unusually red or
unusually blue in the near-infrared are also red or blue, respectively, in
some of the mid-infrared colors. The red objects are best reproduced by
non-equilibrium models with very low sedimentation efficiency ($f_{\rm
sed}\approx 1$) and hence thick condensate cloud decks, and the blue
objects are reproduced by non-equilibrium models with very high sedimentation
efficiency ($f_{\rm sed}\approx 3$--4) and hence thin condensate cloud
decks. 

The models account for most of the observed range in color and absolute magnitude
via variations in the eddy mixing coefficient within the range
$K_{zz}=10^2$--$10^6\,$cm$^2$~s$^{-1}$.
The mixing mechanism in the radiative region of brown dwarf atmospheres is 
presently unknown, so it is not possible to assess whether such a wide range
is reasonable or not.  
The remaining $\lesssim 20$\% discrepancies between observed and modelled colors
and magnitudes can be attributed to
several possible causes, including variations in cloud properties that
are not captured in our one-dimensional cloud model, uncertainties 
in the opacities of CH$_4$ and NH$_3$, the choices of C, N, and O abundances, 
and perhaps variations in the C/O ratio in brown dwarf atmospheres.

Vertical mixing may also be an important consideration for the direct detection 
of giant planets around nearby stars (G04; Marley et al.\ 2006, Hinz et al.\ 2006).  
After the discovery of Gl~229B, Marley et al.\ (1996) suggested that a 
substantial 4--5~$\mu$m flux peak should be a universal feature of giant planets 
and brown dwarfs.  This expectation, combined with a favorable planet/star 
flux ratio, has made the $M$ band a favorite for planet detection (Burrows et al. 2005, Marley et al. 2006).
However,  Figures 4 and 8 show that the $M$ or 4.6~$\mu$m region is fainter -- and the 
$L$ or 3.6~$\mu$m region is brighter -- than the equilibrium chemistry models 
predict, at least for $T_{\rm eff} \gtrsim 700$~K.  Given these and other 
considerations (such as cleaner atmospheric transmission at $L$ compared 
with $M$) the comparative advantage of ground-based searches for young, bright giant planets
at $M$ may be somewhat less than currently expected (see also Marley et al. 2006).
Another implication  of the reduced $M$-band flux is that planetary 
masses (and upper limits) inferred from $M$-band measurements 
using equilibrium chemistry models (Hinz et al.\ 2006) will be too low.

This initial study is a comparison of observed and synthetic broadband 
color--color diagrams of brown dwarfs, and it is limited to identifying and interpreting 
trends in the data.  The model parameters that best match the colors of a specific object 
vary with color, which, unless the models are perfect, limits the level of 
analysis that can be performed.  Being differential in nature, colors can magnify small 
errors in models that otherwise agree well with the data.  For example, the 
$\sim0.3$~mag mismatch between the observed and synthetic [5.73]--[7.78] colors of 
late-T dwarfs (Figures 4(d), 5(d), and 6(d)) betrays the generally good fit of the same models
with the entire SED of the T7.5 dwarf Gl 570D (Saumon et al. 2006).
Because of the haphazard overlap of the photometric and molecular bands and the complex 
variations in brown dwarf spectra, color-color diagrams can be difficult to interpret.
With global trends identified and generally good agreement between models and data,
the detailed study of both typical and anomalous brown dwarfs can be tackled.  Our
analysis of ground-based and {\it Spitzer} photometry and spectra of such objects is 
underway.

\acknowledgments

We are grateful to Malcolm Currie at the Rutherford
Appleton Laboratory, and the Starlink team, for help with the NIRI data
analysis. We are also grateful to Linhua Jiang of the University of Arizona for
help with the IRAC data analysis. Brian Patten of CfA Harvard was very
generous with pre-publication access to IRAC data and in helping with
the IRAC data analysis. 

This work is based in part on observations made with the {\it Spitzer Space
Telescope}, which is operated by the Jet Propulsion Laboratory,
California Institute of Technology under a contract with NASA, 
and on observations made at the Gemini Observatory, 
which is operated by the Association of Universities for Research in
Astronomy Inc. (AURA), under a cooperative agreement with the NSF on behalf of
the Gemini partnership: the National Science Foundation (United States),
the Particle Physics and Astronomy Research Council (United Kingdom),
the National Research Council (Canada), CONICYT (Chile), the Australian
Research Council (Australia), CNPq (Brazil) and CONICET (Argentina).

Support for this work was provided by NASA through an award issued by
JPL/Caltech. 
This work was also supported in part under the auspices of the U.S. Department of
Energy at Los Alamos National Laboratory under Contract W-7405-ENG-36.
MSM acknowledges the support of the NASA Office of Space Sciences.
TRG is supported by the Gemini Observatory, which is operated by AURA
on behalf of the international Gemini partnership listed above.
SKL acknowledges the support of the Gemini Observatory, the Joint Astronomy Centre
and the U.K. Particle Physics and Astronomy Research Council.

Funding for the SDSS and SDSS-II has been provided by the Alfred P.\ Sloan Foundation, 
the Participating Institutions, the National Science Foundation, the U.S.\ Department 
of Energy, NASA, the Japanese Monbukagakusho, 
the Max Planck Society, and the Higher Education Funding Council for England. 
The SDSS Web Site is http://www.sdss.org/. The SDSS is managed by the Astrophysical 
Research Consortium for the participating institutions: 
American Museum of Natural History, Astrophysical Institute Potsdam, University 
of Basel, Cambridge University, Case Western Reserve University, University of Chicago, 
Drexel University, Fermilab, Institute for Advanced Study, Japan Participation Group, 
Johns Hopkins University, Joint Institute for Nuclear Astrophysics, Kavli Institute 
for Particle Astrophysics and Cosmology, Korean Scientist Group, Chinese Academy 
of Sciences (LAMOST), Los Alamos National Laboratory, Max-Planck-Institute for Astronomy 
(MPIA), Max-Planck-Institute for Astrophysics (MPA), New Mexico State University, Ohio 
State University, University of Pittsburgh, University of Portsmouth, Princeton University, 
United States Naval Observatory, and University of Washington.

\vskip0.5cm
\leftline{\bf References}

\noindent\hangindent1em\hangafter1
Ackerman, A. S. \& Marley, M. S. 2001, \apj, 556, 872
\vskip0.1cm
\noindent\hangindent1em\hangafter1
Beichman, C. A., Chester, T. J., Skrutskie, M., Low, F. J. 
         \& Gillett, F. 1998, \pasp, 110, 480
\vskip0.1cm
\noindent\hangindent1em\hangafter1
B\'ezard, B., Lellouch, E., Strobel, D., Maillard J.-P. 
         \& Drossard, P. 2002, Icarus, 159, 95
\vskip0.1cm
\noindent\hangindent1em\hangafter1
Burgasser, A. J. et al. 2002, \apj, 564, 421
\vskip0.1cm
\noindent\hangindent1em\hangafter1
Burgasser, A. J, Marley, M. S., Ackerman, 
A. S., Saumon, D., Lodders, K., Dahn, C. C., Harris, H. C., Kirkpatrick, J. D.
2002, \apj 571, L151
\vskip0.1cm
\noindent\hangindent1em\hangafter1
Burgasser, A. J., Reid, I. N., Leggett, S. K., Kirkpatrick, J. D., Liebert, J., Burrows, A. 2005,
\apjl, 634, 177
\vskip0.1cm
\noindent\hangindent1em\hangafter1
Burgasser, A. J., Burrows, A., 
         Kirkpatrick, J. D. 2006a, \apj, 639, 1095
\vskip0.1cm
\noindent\hangindent1em\hangafter1
Burgasser, A. J., Geballe, T. R., Leggett, S. K., 
         Kirkpatrick, J. Davy \& Golimowski, D. A. 2006b, \apj, 637, 1067
\vskip0.1cm
\noindent\hangindent1em\hangafter1
Burgasser, A. J., Kirkpatrick, J. D., Cruz, K. L., 
         Reid, I. N., Leggett, S. K., Liebert, J., Burrows, A., Brown, M. E. 2006c, \apj, in press, astro-ph/0605577
\vskip0.1cm
\noindent\hangindent1em\hangafter1
Burrows, A., et al. 1997, \apj, 491, 856 
\vskip0.1cm
\noindent\hangindent1em\hangafter1
Burrows, A., Hubbard, W. B., Lunine, J. I., 
         Liebert, J. 2001, Reviews of Modern Physics, 73, 719
\vskip0.1cm
\noindent\hangindent1em\hangafter1
Burrows, A., Sudarsky, D., Hubeny, 
I. 2006, \apj, 640, 1063
\vskip0.1cm
\noindent\hangindent1em\hangafter1
Cavanagh, B., Hirst, P., Jenness, T., Economou, F., 
         Currie, M. J., Todd, S. \& Ryder, S. D., 2003, Astronomical Data Analysis Software 
         and Systems XII, ASP Conference Series, Vol. 295, H. E. Payne, R. I. Jedrzejewski, and R. N. Hook, eds., 237
\vskip0.1cm
\noindent\hangindent1em\hangafter1
Chauvin, G., Lagrange, A.-M., Dumas, C., Zuckerman, B., 
         Mouillet, D., Song, I., Beuzit, J.-L., \& Lowrance, P.\ 2004, \aap, 425, L29 

\noindent\hangindent1em\hangafter1
Chiu, K., Fan, X., Leggett, S. K., 
         Golimowski, D. A., Zheng, W., Geballe, T. R., Schneider, D. P. \& Brinkmann, J. 2006, \aj, 131, 2722, (C06)
\vskip0.1cm
\noindent\hangindent1em\hangafter1
Cushing, M. C. et al. 2006, \apj, 648,614
\vskip0.1cm
\noindent\hangindent1em\hangafter1
Dahn C. C. et al. 2002, \aj, 124, 1170
\vskip0.1cm
\noindent\hangindent1em\hangafter1
Fazio, G. et al., 2004, \apjs, 154, 10
\vskip0.1cm
\noindent\hangindent1em\hangafter1
Fegley, B. Jr. \& Lodders, K.  1994, Icarus, 110, 117
\vskip0.1cm
\noindent\hangindent1em\hangafter1
Fegley, B. Jr. \& Lodders, K.  1996, \apj, 472, L37
\vskip0.1cm
\noindent\hangindent1em\hangafter1
Geballe, T. R. et al. 2002, \apj, 564, 466
\vskip0.1cm
\noindent\hangindent1em\hangafter1
Golimowski D. A. et al. 2004a, \aj, 127, 3516, (G04) 
\vskip0.1cm
\noindent\hangindent1em\hangafter1
Golimowski D. A. et al. 2004b, \aj, 128, 1733
\vskip0.1cm
\noindent\hangindent1em\hangafter1
Griffith, C.A. \& Yelle, R.V. 1999, \apj, 519, L85
\vskip0.1cm
Hinz, P.H., Heinze, A.N., Sivanandam, S., Miller, D.L., Kenworthy, M.A., Brusa, G., 
         Freed, M.  \& Angel, J.R.P. 2006, \apj, in press, astro-ph/0606129
\vskip0.1cm
\noindent\hangindent1em\hangafter1
Hodapp, K. W. et al. 2003, \pasp, 115, 1388
\vskip0.1cm
\noindent\hangindent1em\hangafter1
Houck, J. et al. 2004, \apjs, 154, 18
\vskip0.1cm
\noindent\hangindent1em\hangafter1
Kendall, T. R., Delfosse, X., Mart\'{\i}n, E. L., Forveille, T. 2004, \aap, 416, L17
\vskip0.1cm
\noindent\hangindent1em\hangafter1
Knapp, G. R. et al. 2004, \aj, 127, 3553, (K04)
\vskip0.1cm
\noindent\hangindent1em\hangafter1
Leggett, S. K. et al. 2002, \apj, 564, 452
\vskip0.1cm
\noindent\hangindent1em\hangafter1
Leggett, S. K., Hawarden, T. G., Currie, M. J., Adamson, A. J., 
         Carroll, T. C., Kerr, T. H., Kuhn, O. P., Seigar, M. S., Varricatt, W. P. \& Wold, T. 2003, \mnras, 345, 144
\vskip0.1cm
\noindent\hangindent1em\hangafter1
Liu, M. C. \& Leggett, S. K. 2005, \apj, 634, 616
\vskip0.1cm
\noindent\hangindent1em\hangafter1
Lodders, K. \& Fegley, B., Jr. 2002, Icarus, 393, 424
\vskip0.1cm
\noindent\hangindent1em\hangafter1
Luhman, K. L. et al. 2006, \apj, in press
\vskip0.1cm
\noindent\hangindent1em\hangafter1
Marley, M.S., Fortney, J., Seager, S., Barman, T. 2006, in Protostars and Planets V,
         B. Reipurth, Ed., in press.
\vskip0.1cm
\noindent\hangindent1em\hangafter1
Marley, M.~S., Saumon, D., Guillot, T., Freedman, R.~S., 
         Hubbard, W.~B., Burrows, A., \& Lunine, J.~I.\ 1996, Science, 272, 1919 
\vskip0.1cm
\noindent\hangindent1em\hangafter1
Marley, M.~S., Seager, S., Saumon, D., Lodders, K., Ackerman, A. S., 
         Freedman, R. S. \& Fan, X. 2002, \apj, 568, 335
\vskip0.1cm
\noindent\hangindent1em\hangafter1
Neuh{\"a}user, R., Guenther, E.~W., Wuchterl, G., Mugrauer, M., 
        Bedalov, A., \& Hauschildt, P.~H.\ 2005, \aap, 435, L13 
\vskip0.1cm
\noindent\hangindent1em\hangafter1
Noll, K. S., Geballe, T. R., Leggett, S. K. \& Marley, M. S. 2000, \apj, 541, L75
\vskip0.1cm
\noindent\hangindent1em\hangafter1
Noll, K.S., Geballe, T.R. \& Marley, M.S. 1997, \apj, 489, 87
\vskip0.1cm
\noindent\hangindent1em\hangafter1
Noll, K.S., Knacke, R.F., Geballe, T.R. \& Tokunaga, A.T. 1988, \apj, 324, 1210
\vskip0.1cm
\noindent\hangindent1em\hangafter1
Noll, K.S., \& Larson, H.P. 1991, Icarus, 89, 168
\vskip0.1cm
\noindent\hangindent1em\hangafter1
Patten, B. M., et al. 2006, \apj, in press, astro-ph/0606432
\vskip0.1cm
\noindent\hangindent1em\hangafter1
Reid, I. N. \& Cruz, K. L. 2002, \aj, 123, 466
\vskip0.1cm
\noindent\hangindent1em\hangafter1
Roellig, T. L., et al. 2004, \apjs, 154, 418
\vskip0.1cm
\noindent\hangindent1em\hangafter1
Saumon, D., Geballe, T.~R., Leggett, S.~K.,
         Marley, M.S., Freedman, R.~S., Lodders, K., Fegley, B., Jr. \& Sengupta, S.~K.
         2000, \apj, 541, 374
\vskip0.1cm
\noindent\hangindent1em\hangafter1
Saumon, D., Marley, M.S., Cushing, M.C., Leggett, S.K., Roellig, T.L., Lodders, K.
         \& Freedman, R.S. 2006, \apj, in press, astro-ph/0605563
\vskip0.1cm
\noindent\hangindent1em\hangafter1
Saumon, D., Marley, M. S., Lodders, K. \& Freedman, R. S. 2003, 
         Brown Dwarfs, Proceedings of IAU Symposium 211, ed. E. Mart\'{\i}n.  San Francisco: Astronomical Society 
         of the Pacific, 2003, 345
\vskip0.1cm
\noindent\hangindent1em\hangafter1
Simons, D. A. \& Tokunaga, A. 2002, \pasp, 114, 169
\vskip0.1cm
\noindent\hangindent1em\hangafter1
Skrutskie, M. F. et al. 2006, \aj, 131,  1163
\vskip0.1cm
\noindent\hangindent1em\hangafter1
Stephens, D.~C., Marley, M.~S., Noll, K.~S., \& Chanover, N.\ 2001, \apjl, 556, L97 
\vskip0.1cm
\noindent\hangindent1em\hangafter1
Tokunaga, A. T., Simons, D. A. \& Vacca, W. D. 2002, \pasp, 114, 180
\vskip0.1cm
\noindent\hangindent1em\hangafter1
Toon, O.B., McKay, C.P., Ackerman, T.P. \& Santhanam, K. 1989, JGR 94, 16287
\vskip0.1cm
\noindent\hangindent1em\hangafter1
Vrba, F. J. et al. 2004, \aj, 127, 2948
\vskip0.1cm
\noindent\hangindent1em\hangafter1
Werner, M., et al. 2004, ApJS, 154, 1 
\vskip0.1cm
\noindent\hangindent1em\hangafter1
York, D. G. et al. 2000, \aj, 120, 1579
\clearpage
\begin{deluxetable}{lllllll}
\tabletypesize{\scriptsize}
\tablecaption{Gemini NIRI $L^{\prime}$ and $M^{\prime}$ Photometry (MKO System)}
\tablewidth{430pt}
\tablehead{
\colhead{Name} &  \colhead{Type\tablenotemark{1}} &
\colhead{$L^{\prime}$(error)} & \colhead{Date} &
\colhead{$M^{\prime}$(error)} & \colhead{Date} & Notes \\

\colhead{} & \colhead{} & \colhead{(mag)} & \colhead{(YYMMDD)} & \colhead{(mag)} & \colhead{(YYMMDD)} & \\
}
\startdata
SDSS J000013.54+255418.6 & T4.5  & 13.03(0.03) & 040826 & 13.28(0.10)  & 040801 & \nodata \\
2MASS J00345157+0523050  &   T6.5  & 13.30(0.04) & 040827 & 12.66(0.10)  & 040827 & \nodata \\
2MASS J02431371-2453298   &     T6   &  \nodata &  \nodata & 13.27(0.16)  &  050128 &\nodata \\
SDSS J080531.80+481233 & L9.5 & 12.34(0.04) & 041230 & 12.81(0.07) & 041230,050115 & 2,3\\
SDSS J083008.12+482847.4  &  L9  & \nodata & \nodata & 12.75(0.15)  &  040405 & \nodata \\
SDSS J085758.45+570851.4 &  L8  & \nodata  & \nodata & 11.55(0.10)  &  040405 & 4,5 \\
SDSS J111010.01+011613.1 &    T5.5  & 13.89(0.05) & 050115 & \nodata  & \nodata  & 4 \\
SDSS J133148.90-011651.4   &    L8 & 12.73(0.03) & 050115 & \nodata    & \nodata  & 2\\
2MASS J15031961+2525196   &  T5  & \nodata &  \nodata & 11.92(0.04)  &  040509 & 6\\
2MASS J15530228+1532369AB  &  T7    &   13.83(0.05) &  050115 &  13.29(0.15)  &  040406 & 7 \\
2MASS J16322911+1904407  &   L7.5  & \nodata & \nodata & 13.07(0.15)  &  040530 & \nodata \\
2MASS J22443167+2043433 &  L7.5 &  \nodata  &  \nodata  &11.90(0.03)  &  040801 & 4\\

\enddata

\tablenotetext{1}{~Spectral types are based on near-infrared classification schemes for L dwarfs
(Geballe et al.\ 2002) and T dwarfs (Burgasser et al.\ 2006b). }
\tablenotetext{2}{~Near-infrared colors are unusually blue for this spectral type.}
\tablenotetext{3}{~G04 reported $L^{\prime}=12.31\pm 0.05$. }
\tablenotetext{4}{~Near-infrared colors are unusually red for this spectral type.}
\tablenotetext{5}{~Leggett et al.\ (2002) reported $M^{\prime}=11.50\pm 0.05$. }
\tablenotetext{6}{~G04 reported $M^{\prime}=12.25\pm 0.15$. }
\tablenotetext{7}{~T7 type applies to unresolved pair.  Burgasser et al.\ (2006c) report T6.5+T7.}

\end{deluxetable}

\begin{deluxetable}{cccc}
\tabletypesize{\scriptsize}
\tablecaption{Filter Bandpasses at Operational Temperatures}
\tablewidth{250pt}
\tablehead{
\colhead{Filter} & \colhead{Nominal} & \colhead{50\% Cut-}
& \colhead{50\% Cut-}\\
\colhead{Name} & \colhead{Wavelength ($\mu$m)} & \colhead{On ($\mu$m)} &
\colhead{Off ($\mu$m)}\\
}
\startdata
$L^{\prime}_{\rm MKO}$ &  3.75 & 3.43  & 4.11 \\
$M^{\prime}_{\rm MKO}$  &  4.70 & 4.57 &  4.80 \\
IRAC channel 1 & 3.55 & 3.18 &  3.92 \\
IRAC channel 2 & 4.49 & 4.00 &  5.01 \\
IRAC channel 3 & 5.73 & 5.02  & 6.43 \\
IRAC channel 4 & 7.87 & 6.45  & 9.33 \\

\enddata
\end{deluxetable}

\clearpage
\voffset 1.8truein
\begin{deluxetable}{llrrrcrrrcrrrcrrrcl}
\rotate
\tabletypesize{\scriptsize}
\tablecaption{$Spitzer$ IRAC Photometry}
\tablewidth{665pt}
\tablehead{
\colhead{Name} & \colhead{Type\tablenotemark{1}} & \multicolumn{3}{c}{Channel 1} & \colhead{} &  \multicolumn{3}{c}{Channel 2} & \colhead{}& \multicolumn{3}{c}{Channel 3} & \colhead{}& \multicolumn{3}{c}{Channel 4}  & \colhead{Date} & Notes \\
\cline{3-5}\cline{7-9}\cline{11-13}\cline{15-17}
 &  & \colhead{mJy} & \colhead{[3.55]} & \colhead{\% err} &  \colhead{} & \colhead{mJy} & \colhead{[4.49]} & \colhead{\% err} &   \colhead{} & \colhead{mJy} & \colhead{[5.73]} & \colhead{\% err} & \colhead{} & \colhead{mJy} & \colhead{[7.87]} & \colhead{\% err} & \colhead{(YYMMDD)} & \\
}
\startdata
SDSS J000013.54+255418.6 & T4.5  & 0.91 & 13.72 & 1.0 & & 1.06 & 13.07 & 2.0 & & 1.09 & 12.56 & 10.0 & & 0.64 & 12.50 & 1.0 & 050725 & 2 \\
SDSS J075840.33+324723.4 & T2 & 3.05 & 12.41 & 0.4 & & 2.52 & 12.13 & 0.5 & &2.37 & 11.71 & 1.8 & & 1.89 & 11.33 & 4.0 & 051023 & \nodata\\
SDSS J080531.80+481233 & L9.5 & 2.98 & 12.44 & 0.5 & & 1.92 & 12.43 & 0.5 & & 1.36 & 12.32 & 1.1 & & 0.93 & 12.10 & 1.8 & 051022 & 3\\
SDSS J105213.51+442255.7 & T0.5 & 1.43 & 13.23 & 0.1 & & 1.03 & 13.10 & 0.2 & & 0.98 & 12.67 & 0.5 &&  0.73 & 12.36 & 0.8 & 051127 & \nodata \\
SDSS J115553.86+055957.5 & L7.5 & 2.07 & 12.83 & 0.2 & & 1.33 & 12.83 & 0.3 & & 1.15 & 12.50 & 1.0&  & 0.80 & 12.26 & 2.0 & 051229 & \nodata\\
SDSS J120747.17+024424.8 & T0 & 2.01 & 12.86 & 0.2 & & 1.46 & 12.73 & 0.3 & & 1.24 & 12.42 & 1.5 & & 0.82 & 12.23 & 5.0 & 060706 &  \nodata\\
SDSS J151643.01+305344.4 & T0.5 & 1.01 & 13.61 & 0.5 & & 0.79 & 13.39 & 1.0 & & 0.80 & 12.89 & 1.0 & & 0.66 & 12.47 & 2.0 & 050724 & 4\\
SDSS J152039.82+354619.8 & T0 & 1.87 & 12.94 & 0.5 & & 1.25 & 12.89 & 0.2 & & 1.16 & 12.49 & 1.2 & & 0.80 & 12.26 & 0.8 & 050726 & \nodata\\
2MASS J22443167+2043433  &  L7.5 & 3.22 & 12.35 & 0.4 & & 2.58 & 12.11 & 0.5 & & 2.65 & 11.59 & 0.9 &  &1.83 & 11.36 & 1.5 & 050723 & 5
\enddata
\tablenotetext{1}{~Spectral types are based on near-infrared classification schemes for L dwarfs
(Geballe et al.\ 2002) and T dwarfs (Burgasser et al.\ 2006b). }
\tablenotetext{2}{~Nearby bright star caused more uncertain photometry.  Sky values were interpolated between the diffraction spikes of the bright star.}
\tablenotetext{3}{~Near-infrared colors are unusually blue for this spectral type. } 
\tablenotetext{4}{~Near-infrared colors are unusually red for this spectral type. }
\tablenotetext{5}{~Near-infrared colors are extremely red for this spectral type.}
\end{deluxetable}

\clearpage

\end{document}